\documentclass[twocolumn]{aastex631}

\received{\today}
\revised{--}
\accepted{--}

\usepackage[utf8]{inputenc}
\usepackage{graphicx}
\usepackage{bm}
\usepackage{color, soul}      
\usepackage{units}
\usepackage{amsmath}
\usepackage{epsfig}
\usepackage{amssymb}
\usepackage{mathrsfs}
\usepackage{comment}

\allowdisplaybreaks
\def\be{\begin{align}}
\def\ee{\end{align}}
\def\bea{\begin{eqnarray}}
\def\eea{\end{eqnarray}}
\def\ban{\begin{eqnarray*}}
\def\ean{\end{eqnarray*}}
\def\bd{\begin{displaymath}}
\def\ed{\end{displaymath}}

\def\bc{\begin{center}}
\def\ec{\end{center}}

\def\ba{\begin{array}}
\def\ea{\end{array}}

\shorttitle{Formation of hammerhead-like distributions}
\shortauthors{L\'opez et al.}

\begin{document}

\title{Hybrid simulations of the proton beam instabilities in the young solar wind. \\ The formation of hammerhead-like distributions}
\author[0000-0003-3223-1498]{R. A. L\'opez}
\email{rodrigo.lopez@cchen.cl}
\affiliation{Research Center in the intersection of Plasma Physics, Matter, and Complexity ($P^2 mc$), Comisi\'on Chilena de Energ\'{\i}a Nuclear, Casilla 188-D, Santiago, Chile}
\affiliation{Departamento de Ciencias F\'{\i}sicas, Facultad de Ciencias Exactas, Universidad Andres Bello, Sazi\'e 2212, Santiago 8370136, Chile}
\author[0000-0003-0465-598X]{Shaaban M. Shaaban}
\affiliation{Department of Physics and Materials Sciences, College of Arts and Sciences, Qatar University, 2713 Doha, Qatar.}
\author[0000-0002-8508-5466]{M. Lazar}
\affiliation{Centre for mathematical Plasma Astrophysics, Department of Mathematics, KU Leuven, Celestijnenlaan 200B, 3001 Leuven, Belgium}
\affiliation{Theoretical Physics IV, Faculty for Physics and Astronomy, Ruhr-University Bochum, D-44780 Bochum, Germany.}
\author[0000-0001-5079-7941]{L. Pezzini}
\affiliation{Centre for mathematical Plasma Astrophysics, Department of Mathematics, KU Leuven, Celestijnenlaan 200B, 3001 Leuven, Belgium}
\affiliation{Solar-Terrestrial Centre of Excellence--SIDC, Royal Observatory of Belgium, 1180 Brussels, Belgium}
\author[0000-0002-1743-0651]{S. Poedts}
\affiliation{Centre for mathematical Plasma Astrophysics, Department of Mathematics, KU Leuven, Celestijnenlaan 200B, 3001 Leuven, Belgium}
\affiliation{Institute of Physics, University of Maria Curie-Sk{\l}odowska, Pl.\ M.\ Curie-Sk{\l}odowska 5, 20-031 Lublin, Poland.}
\author[0000-0002-9151-5127]{H. Fichtner}
\affiliation{Theoretical Physics IV, Faculty for Physics and Astronomy, Ruhr-University Bochum, D-44780 Bochum, Germany.}
\author[0000-0001-8134-3790]{P. H. Yoon}
\affiliation{Institute for Physical Science and Technology,
University of Maryland, College Park, MD 20742-2431, USA}
\begin{abstract}
Parker Solar Probe (PSP) observations in the young solar wind reveal new properties of both plasma particle velocity distributions (VDs) and associated electromagnetic (EM) wave fluctuations.
The quasilinear (QL) kinetic theory of plasma wave instabilities has recently shown that new hammerhead (HH) proton distributions can be generated by the relaxation of proton beams through the instabilities of right-handed (RH) polarized waves. 
Such RH waves have indeed been reported in association with HH distributions.
In this paper, new results from hybrid simulations of proton-beam-plasma systems with properties typical of those observed to excite EM-RH wave instabilities are presented. 
From the long-term evolution of these systems, it is found that beam relaxation is driven by instabilities and growing wave fluctuations, leading to HH-type features in the velocity distributions.
The production of these features, as well as their prominence, depends on the magnetic power of the waves generated by the instabilities and, therefore, implicitly on the available free energy, quantified by the plasma beta parameter and the relative beam drift.
The simulation results capture the self-consistent evolution of the instabilities and their nonlinear development. Linear theory, together with simulations, helps identify the nature of the unstable modes and the plasma conditions under which they arise. The good agreement with quasi-linear (QL) theory further indicates that it can serve as a computationally efficient complementary framework for interpreting the associated wave-particle interactions.

\end{abstract}

\keywords{Solar wind (1534); Space plasmas (1544); Plasma physics (2089)}

\section{Introduction} \label{s1}

Recent in-situ measurements in the young solar wind reveal the existence of low-frequency waves with right-handed (RH) circular polarization in association with proton or ion beams and occasionally the so-called hammer-head (HH) profiles in the velocity distributions (VDs)\citep{Klein2021+ApJ, Verniero-2020, Verniero-etal-2022, Ofman_2022, Phan-etal-2022, BharatiDas2026}. 
Proton and heavy-ion beams can have significant implications for various processes in space plasmas, especially in the solar wind and (inter)planetary shocks. 
Recent observations by Parker Solar Probe (PSP) and Solar Orbiter (SolO) support the hypothesis that physical processes adjacent to magnetic reconnection in current sheets can produce proton beams, at lower near-thermal energies across the heliospheric current sheet, and at suprathermal energies in solar flares, coronal mass ejections, and shocks \citep{Matteini2013,Desai-Giacalone-2016,Lavraud-etal-2021, Phan-etal-2022,Fargette2026}. In this context, \cite{Louarn_2021AA} further discuss the correlation between the orientation of magnetic flux tubes and the development of proton beams, finding a significant beam enhancement in the vicinity of discontinuities, including magnetic switchbacks.
In existing theories, the interactions between ions (including protons), plasma waves, and turbulence play a central role in the formation of these beams \citep{Tu-etal-2002, Araneda-etal-2008, Pierrard-Voitenko-2010, Voitenko-Pierrard-2015,Pezzini2026}.
Furthermore, suprathermal populations, including ion/proton beams, are believed to be seed populations that, for example, are accelerated in interplanetary shocks, producing solar energetic particles \citep{Lario2019+ApJ, Lario2022+ApJ}.

The enhanced RH waves observed by PSP suggest that, on the one hand, these waves may result from proton-beam instabilities, and, on the other hand, the same fluctuations can contribute to the relaxation of the proton beams.
The first hypothesis is supported by the theory of linear wave dispersion and stability \citep{Klein2021+ApJ, Shaaban2024+AA}, and by preliminary fully kinetic simulations \citep{Pezzini-etal-2024}, which used parameters inspired by the observations in \cite{Klein2021+ApJ}.
Regarding the RH-polarized fluctuations generated by the ion-ion (firehose-like) instability of proton beams, the resulting wave intensity increases if the beam velocity increases, and QL theory proves a major back effect of the enhanced fluctuations on the beam, contributing to its relaxation and even to the formation of HH features \citep{Shaaban2024+AA}.

In this paper, we present new results from hybrid simulations (protons as particles and electrons as a neutralizing fluid), using typical parameterizations for proton beams, as observed by PSP \citep{Klein2021+ApJ, Verniero-2020, Verniero-etal-2022}, see Tables~\ref{t1} and \ref{t2} below.
The aforementioned PIC simulations \citep{Pezzini-etal-2024}, restricted to the observed Alfvénic beam speed, showed beam relaxation but without strong evidence of HH features.
Here, we confirm predictions from QL theory \citep{Shaaban2024+AA} on the formation of HH distributions, showing that these effects depend on the level of wave fluctuations produced by the instability and implicitly on the (initial) amount of free kinetic energy in the non-thermal distribution.

The manuscript is structured as follows. In Section~\ref{s2}, we introduce the VD models for the proton core and beam populations, counter-drifting, and describe each of them using a bi-Maxwellian VD.
After motivating the parameterizations used, the unstable solutions of RH transverse waves derived from linear kinetic theory are presented.
In Section~\ref{s3}, we examine in detail the temporal evolution of RH electromagnetic fluctuations and their feedback on the VD, as obtained from numerical simulations. The main findings are summarized in Section~\ref{s4}.

%
\section{Dispersion and stability formalism of proton-beam plasmas}\label{s2}
%
\subsection{\bf Models of proton-beam VDs}
%
Motivated by recent PSP observations \citep{Klein2021+ApJ,Verniero-etal-2022}, we consider a collisionless and quasi-neutral plasma of electrons and protons, with the latter exhibiting a dual structure with a dense core (subscript ``$c$") and a dilute, yet hotter beam component (subscript ``$b$"). 
Thus, the total proton VD is given by 
\begin{align}\label{e1}
    f_p=\frac{n_c}{n_p} f_c+\frac{n_b}{n_p} f_b,
\end{align}
where $n_c$ and $n_b$ are the number densities of the core and beam proton populations, respectively, and $n_p=n_c+n_b$ is the total number density of protons. 
Initially, the core and beam populations have a drifting bi-Maxwellian distribution
\begin{align}\label{e2}
f_{j}\left(v_{\parallel }, v_{\perp }\right) =& \; \frac{1}{\pi^{3/2} \theta_{\parallel j}^{2}\theta_{\perp j}}\exp \left(-\frac{(v_{\parallel}-U_j)^{2}}{\theta_{\parallel j}^{2}}
-\frac{v_{\perp }^{2}}{\theta_{\perp j}^{2}}\right),   
\end{align}
where $j$ is the species index, $U_j$ is the drift velocity, and $\theta_{\parallel j}$ ($\theta_{\perp j}$) is the parallel (perpendicular) thermal speed, $\theta_{\parallel j}=(2k_BT_{\parallel j}/m_p)^{1/2}$, with $m_p$ the proton mass, $T_{\parallel}$ ($T_{\perp}$) the temperature in the direction parallel (perpendicular) to the background magnetic field, and $k_B$ the Boltzmann constant. The current neutrality is maintained by enforcing $n_c U_{c} = -n_b U_{b}$.

Given that the primary focus of this study is the investigation of RH electromagnetic fluctuations via hybrid simulations, the influence of electrons (subscript $e$) is minimized by treating them as a massless fluid, while in the linear analysis, they are assumed to be isotropic and Maxwellian, as
\begin{align}\label{eq4}
f_{e}\left( v\right) =&\frac{1}{\pi^{3/2} \theta_{e}^{3}}\exp \left(-\frac{v^{2}} {\theta_e^{2}}\right),   
\end{align}
where the electron thermal velocity $\theta_e=(2k_BT_e/m_e)^{1/2}$, with $m_e$ the electron mass.

%
\begin{table}
    \centering
     \caption{Plasma parameters from \cite{Klein2021+ApJ} used in simulations (Case~1). Other plasma parameters are $n_b/n_c=0.157$, $T_{\parallel,b}/T_{\parallel,c}=2.465$, $\theta_{\parallel,c}/c=1.861\times10^{-4}.$}
    \begin{tabular}{c|cccc}
    \hline\hline
        Population & $n_j/n_e$ & $\beta_{\parallel,j}$ & $T_{\perp,j}/T_{\parallel,j}$ & $U_{j}/v_A$\\\hline
        Core (c) & $0.864$ & $0.41$ & $0.77$ & $-0.195$ \\
        Beam (b) & $0.1357$ & $0.158$ & $0.62$ & $1.247$ \\
        Electrons (e)& $1.0$ & $0.474$ & $1.0$ & $0.0$\\
        \hline
    \end{tabular}
    \label{t1}
\end{table}
%

%
\begin{table}
    \centering
     \caption{Plasma parameters estimated from \cite{Verniero-etal-2022} used in simulations (Case~2). Other plasma parameters are $T_{\parallel,b}/T_{\parallel,c}=2.465$, $\theta_{\parallel,c}/c=1.861\times10^{-4}.$}
    \begin{tabular}{c|cccc}
    \hline\hline
        Population & $n_j/n_e$ & $\beta_{\parallel,j}$ & $T_{\perp,j}/T_{\parallel,j}$ & $U_{j}/v_A$\\\hline
        Core (c) & $0.95$ & $1.0$ & $1.0$ & $-0.236$ \\
        Beam (b) & $0.05$ & $0.13$ & $1.0$ & $4.5$ \\
        Electrons (e)& $1.0$ & $1.05$ & $1.0$ & $0.0$\\
        \hline
    \end{tabular}
    \label{t2}
\end{table}
%

%
\subsection{\bf Linear theory}
\begin{figure}[t!]
    \centering
    \includegraphics[width=0.48\textwidth]{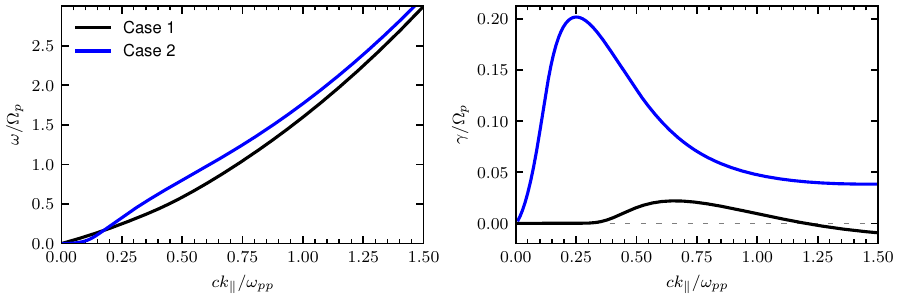}
    \includegraphics[width=0.47\textwidth]{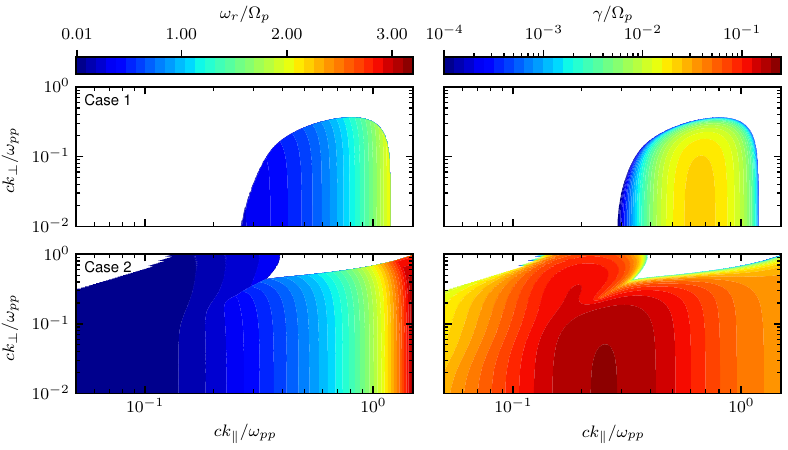}
     \caption{Unstable solutions from linear theory: normalized wave-frequency (left) and growth-rates (right), as functions of $k_\parallel$ (upper) and both $k_\parallel$ and $k_\perp$ (lower). The cases correspond to Klein's parameters in Table~\ref{t1} and Verniero's parameters in Table~\ref{t2}.}
    \label{fig:linear}
\end{figure}

%
Using the DIS-K solver~\citep{Lopez-etal-2021,Lopez2023}, we solve the dispersion and stability of the RH modes for the plasma parameters in Table~\ref{t1}, indicated by the PSP observations in \citet{Klein2021+ApJ}, and Table~\ref{t2}, inspired by \citet{Verniero-etal-2022}. Hereafter, these are referred to as Case~1 and Case~2, respectively. In the tables, $\beta_{\parallel,j}=8\pi n_jT_{\parallel,j}/B^2$ is the parallel plasma beta of species $j$, $\beta_{\parallel b}
= \beta_{\parallel c}
\left( n_b/n_c \right)
\left( T_{\parallel b}/T_{\parallel c} \right)$,  $v_A=B/\sqrt{4\pi n_pm_p}$ is the Alfv\'en speed for the protons, and $c$ is the speed of light.
Figure~\ref{fig:linear} upper panels display the unstable solutions for parallel propagation ($\theta=0^\circ$) for both models: Case 1 (black) and Case 2 (blue), as listed in Table~\ref{t1} and in Table~\ref{t2}, respectively.  Moreover, the lower panels display the full spectrum of unstable RH modes, with color-coded frequencies (left) and growth rates (right), in the $k_\parallel$--$k_\perp$ plane. 

Although the real frequencies in both models are similar, the growth rates of the RH waves differ significantly. Recent results by \citet{Shaaban2025_ApJ} have shown that increasing the plasma beta enhances the instability growth rate. For instance, increasing $\beta_{\parallel c}$ from 0.41 to 1.0 leads to approximately a two times higher growth rate (see Figure~2 therein). In the present study, the growth rates are computed not only for higher $\beta_{\parallel c}$ values but also for higher drift velocity and lower beam density in Case~2 compared to Case~1. The resulting growth rate in Case~2 is about two orders of magnitude higher than that in Case~1, indicating that increasing the drift velocity has a cumulative and significant stimulating effect on the instability, even for dilute beams. The lower panels of Figure~\ref{fig:linear} show that the maximum growth rates for both cases occur in directions quasi-parallel to the background magnetic field. The peak (maximum) of the growth rate shifts toward lower wavenumbers for stronger instabilities, whereas the instability diminishes at higher propagation angles. An interesting feature observed in the growth rates of Case~2 is the hump-shaped structure that appears at higher angles and even lower wavenumbers. This secondary instability occurs around $\theta=60^\circ$. Although this feature is smaller than the instability's main peak, it remains more pronounced than the maximum growth rate in Case~1, suggesting important implications for the two-dimensional simulations discussed in the next section.

\section{\bf Hybrid simulations} \label{s3}

\subsection{\bf Simulation setup}

To study the temporal evolution of these instabilities, we employ explicit hybrid simulations, modeling protons as described above, while electrons are treated as a massless, charge-neutralizing fluid with a constant temperature.
We use the code Hybrid-VPIC~\citep{Bowers2008,Le2021,Le2023}, a high-performance open-source hybrid PIC code available at \href{https://github.com/lanl/vpic-kokkos/tree/hybridVPIC}{https://github.com/lanl/vpic-kokkos/tree/hybridVPIC}. 
The 2D setup in the $x$--$y$ plane consists of $N_x\times N_y=1024\times1024$ cells in a box of length $L_x=L_y=256\,v_A/\Omega_{p}$, with $1024$ particles-per-cell per component (core and beam protons), and where $\Omega_p=eB/(m_pc)$ is the proton gyro-frequency. The time-step used is $\Delta t=0.01\,/\Omega_p$. Given the disparity in growth rates between the two cases, as shown before, and to save computational resources, the simulations run until  $T_\text{max}=655.36\,/\Omega_p$ for case 1 and $T_\text{max}=327.68\,/\Omega_p$ for case 2.
The constant background magnetic field is oriented along the $x$--direction.
%
%
\begin{center}
    \begin{figure}
        \centering
        \includegraphics[width=0.98\linewidth]{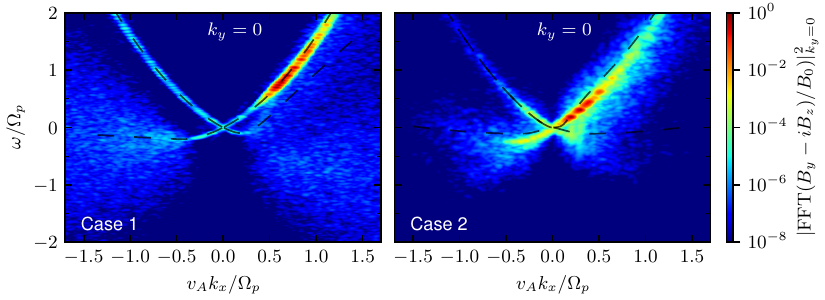}        
        \includegraphics[width=0.98\linewidth]{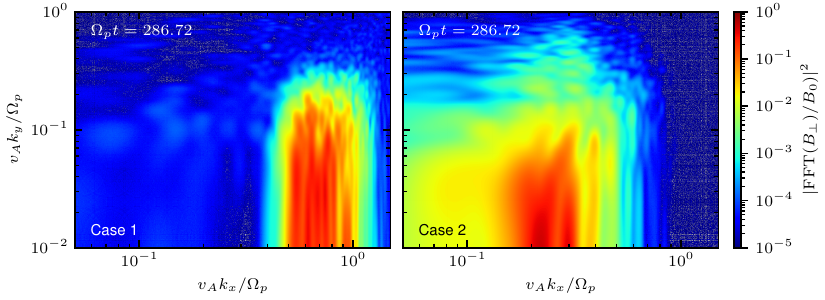}    
        \caption{(top panels) Normalized power spectra for the transverse magnetic field obtained from the FFT of $B_y-iB_z$ at $k_y=0$. RH waves are shown in $\omega>0$ and LH waves in $\omega<0$. Black dashed lines correspond to the linear theory solutions at $\theta=0^\circ$. (bottom panels) Snapshot of the normalized power spectra for the transverse magnetic field.}
        \label{fig:spectra}
    \end{figure}
\end{center}

\subsection{\bf RH instabilities}
%
Beyond the linear analysis presented in Figure~\ref{fig:linear}, this section investigates the enhanced fluctuations associated with the RH unstable modes driven by initially ($t=0$) counter-drifting anisotropic proton populations and then their back reaction on the VDs. We set the simulation using the plasma parameters in Tables \ref{t1} and \ref{t2} as reported by the PSP observations in \citet{Klein2021+ApJ} and \citet{Verniero-etal-2022}.

As predicted by linear theory, the more intense growth rates in Case~2 also span a wider range of wave numbers, most likely stimulating wave-particle interactions. We can then expect them to lead to great changes in the VDs and their moments. In Fig.~\ref{fig:spectra} (top panels), we show the dispersion relation $\omega$ vs. $k_x$, at $k_y=0$, obtained by the FFT of $B_y-iB_z$. This allows us to separate the right-handed contribution ($\omega>0$) from the left-handed one ($\omega< 0$) \citep{Saeed2017}. It is clear that in both cases we have a right-handed instability, as predicted by linear theory in Fig.~\ref{fig:linear}. In Fig.~\ref{fig:spectra} bottom panels we show a snapshot of the normalized power spectra of the transverse fluctuations in the $k_x$ vs. $k_y$ space, showing a remarkable similarity with Fig.~\ref{fig:linear} (bottom), with Case 1 showing the power concentrated at a narrow band in $k_x$, and Case 2 spanning for a wider range in $k_x$ and showing secondary features at higher values of $k_y$. 

\begin{figure}[t!] \centering 
  \includegraphics[width=0.8\linewidth]{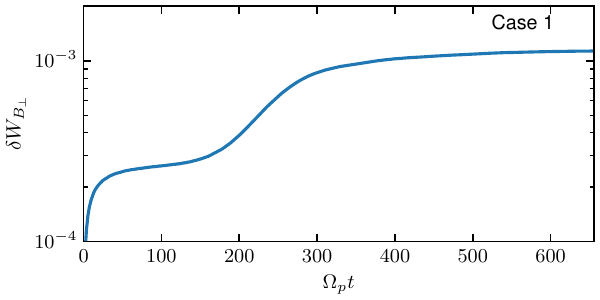}   
  \includegraphics[width=0.8\linewidth]{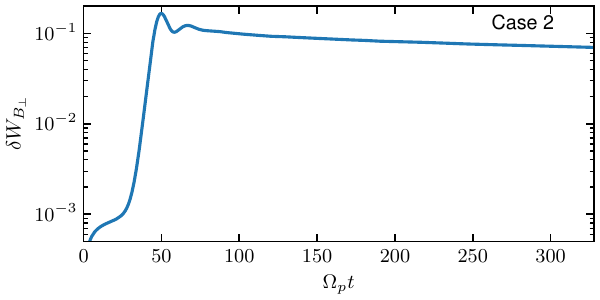}       
  \caption{Rate of change of the transverse magnetic energy density, $[W_{B_{\perp}}-W_{B_{\perp}}(0)]/W_{B_{\perp}}(0)$, for cases 1 and 2.}
        \label{fig:energy}
\end{figure}

The results in Figure~\ref{fig:energy} describe the temporal evolution of the rate of change of the transverse magnetic energy density of the simulation, $\delta W_{B_\perp} (t) =[W_{B_\perp}(t)-W_{B_\perp}(0)]/W_{B_\perp}(0)$, for cases 1 (top panel) and 2 (bottom panel). Here $\perp$ refers to the plane perpendicular to the background magnetic field. It is evident that the enhanced fluctuations in Case~2 develop more rapidly and steeply, reaching a higher saturation level—approximately two orders of magnitude greater than that in Case~1. The $\delta W_{B_\perp}$ attains its peak at a very short time, $t \approx 50/\Omega_p$, occurring even earlier than in Case~1, which still requires time to exhibit considerable growth, hence the reason to run Case 1 for a longer time. These results are consistent with the predictions of linear theory shown in Figure~\ref{fig:linear}.

%

%
\begin{center}
    \begin{figure*}\centering
\includegraphics[width=0.95\linewidth]{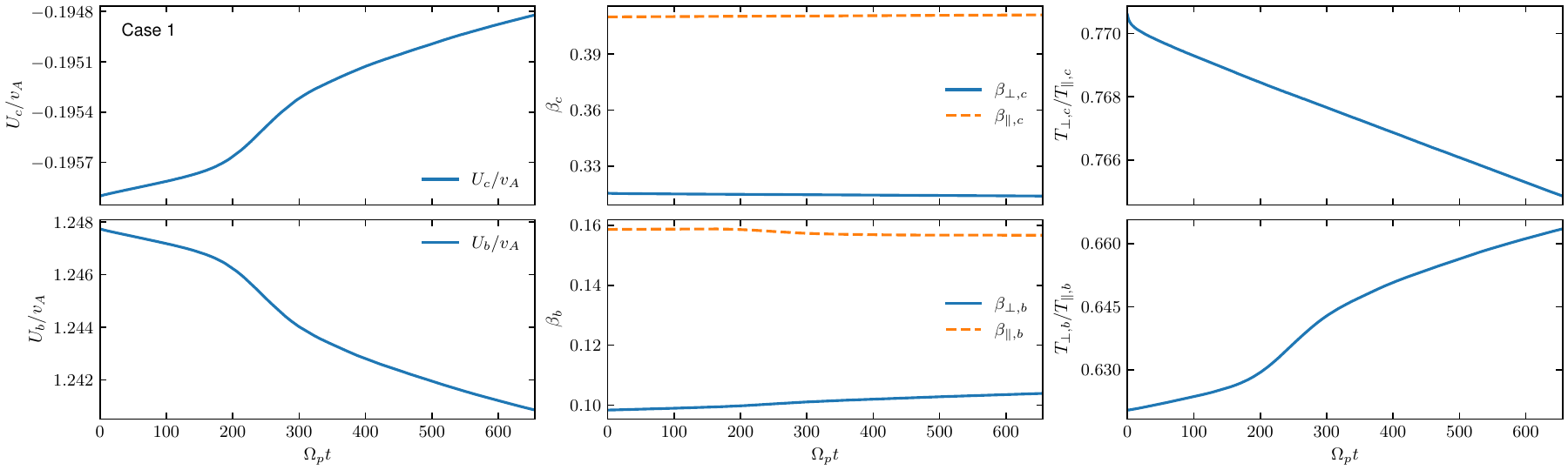}
\includegraphics[width=0.95\linewidth]{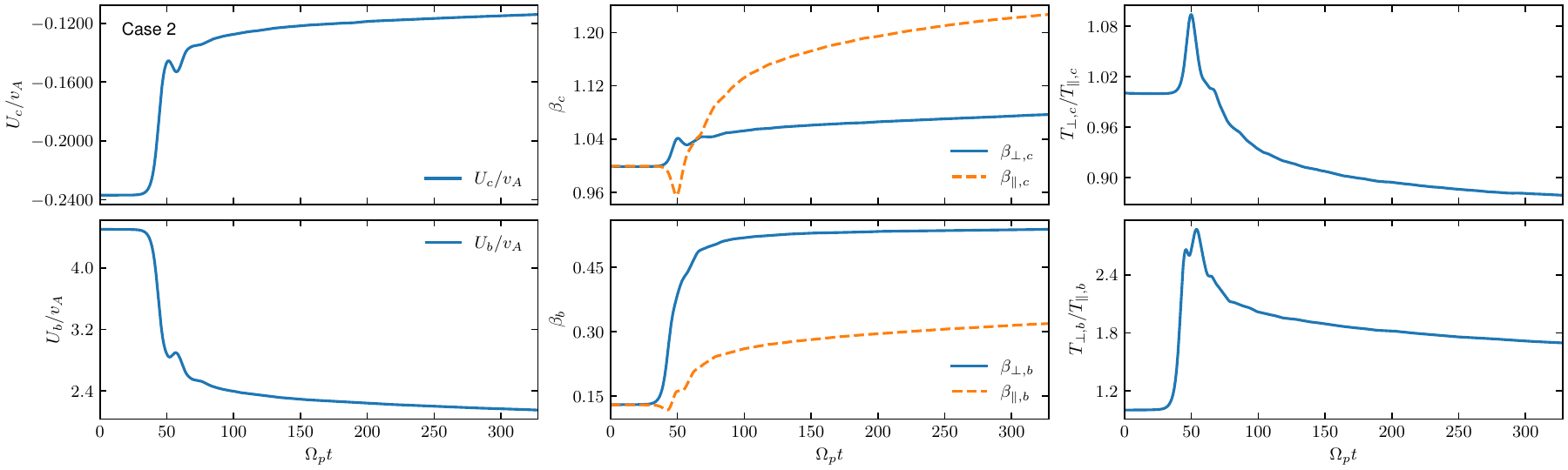}
        \caption{Moments of the distribution for both species, proton core (top) and beam (bottom). The columns show the drift velocity, plasma beta, and temperature anisotropy.}
        \label{fig:moments}
    \end{figure*}
\end{center}

Figure~\ref{fig:moments} shows the temporal evolution of the moments for both species: drift velocities, $U/v_A$, in the left panels, parallel and perpendicular plasma beta parameters ($\beta_{j,\parallel,\perp}=8\pi n_j T_{j,\parallel,\perp}/B_0^2$) in the middle panels, and temperature anisotropies $T_\perp / T_\parallel$ in the right panels, for Case~1 (first and second rows) and Case~2 (third and fourth rows). 

Starting with Case 1, a direct consequence of the enhanced fluctuations of the RH modes is the relaxation of the main driver of the instability, namely the counter-drifting velocities of the beam, $U_{b}(0) = 1.247\,v_A$, and the core, $U_{c}(0) = -0.195\,v_A$, which relaxed to approximately $1.241\,v_A$ and $-0.1948\,v_A$, respectively, at the final stage (top-left and middle-first left panels). However, the underlying wave–particle interaction is more complex, as the drift relaxation is accompanied by parallel cooling and perpendicular heating processes that modify the particle distribution, as reflected in the beam plasma beta parameters, $\beta_{\parallel,\perp} \propto T_{\parallel,\perp}$, shown in the top-middle panel. These cooling and heating processes are negligible for the core populations (middle panel of the second row). An alternative visualization of these cooling and heating processes is obtained by considering the temperature anisotropy, defined as $A = \beta_\perp / \beta_\parallel$, for the beam and core populations, as shown in the left panels of the first and second rows, respectively. The initially anisotropic beam, with $A_b(0) = 0.62$, relaxes to a lower anisotropy of $A_b \approx 0.66$ at the final stage, corresponding to an approximate $6.5\%$ reduction in anisotropy. In contrast, the core anisotropy remains nearly unchanged, with variations on the order of $10^{-3}$. It is worth noting that the initial beam anisotropy cannot trigger the firehose anisotropy-driven instability for the used plasma beta, $\beta_b = 0.158 < 1$. However, the presence of this anisotropy plays a critical role in stimulating ion-ion (firehose-like) instability \citep{Shaaban2025_ApJ}. 

As discussed earlier in Figs.~\ref{fig:spectra} and \ref{fig:energy}, the enhanced fluctuations in Case~2 develop much more rapidly, exhibit steeper growth, and reach significantly higher levels after saturation than those in Case~1. A direct consequence of these stronger fluctuations is an intensified wave–particle interaction, which drives a more pronounced relaxation of the initial unstable plasma parameters. In particular, the counter-streaming velocities are substantially reduced. For example, the initial beam drift speed, $U(0)/v_A = 4.5$, relaxes to approximately $2.1$ at the final stage, corresponding to a decrease of nearly $53\%$ from its initial value (left-bottom panel). This pronounced relaxation of the drift velocity is accompanied by significant heating in both directions relative to the background magnetic field, as indicated by the plasma betas (third and fourth rows, middle panels). For the beam population, heating is more pronounced in the perpendicular direction (fourth row, middle panel), whereas for the core population, it is more pronounced in the parallel direction (third row, middle panel). As a consequence, the initially isotropic beam develops a strong perpendicular temperature anisotropy, reaching a peak value of $A_b \approx 2.8$ (an increase of about $180\%$ relative to its initial value) before gradually decreasing to $A_b \approx 1.7$ (approximately $70\%$ above its initial value) at the final stage (fourth-row right panel). In contrast, the initially isotropic core population exhibits a moderate parallel temperature anisotropy, attaining $A_c \approx 0.9$ at the end of the evolution (third-row right panel).

\subsection{\bf Relaxation of proton VDs}
%
\begin{center}
    \begin{figure*}\centering
\includegraphics[width=0.48\linewidth]{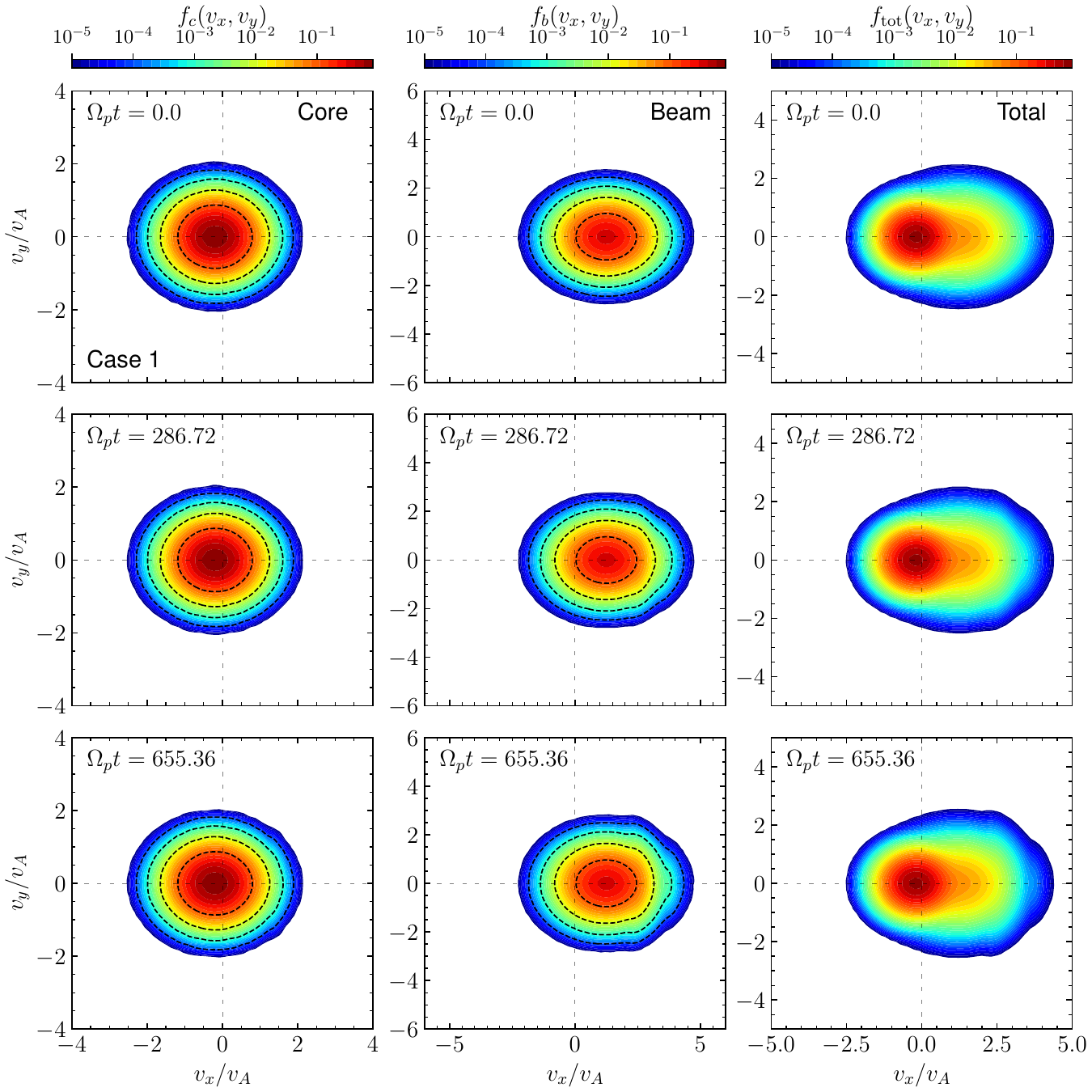} \hspace{0.02\linewidth}
\includegraphics[width=0.48\linewidth]{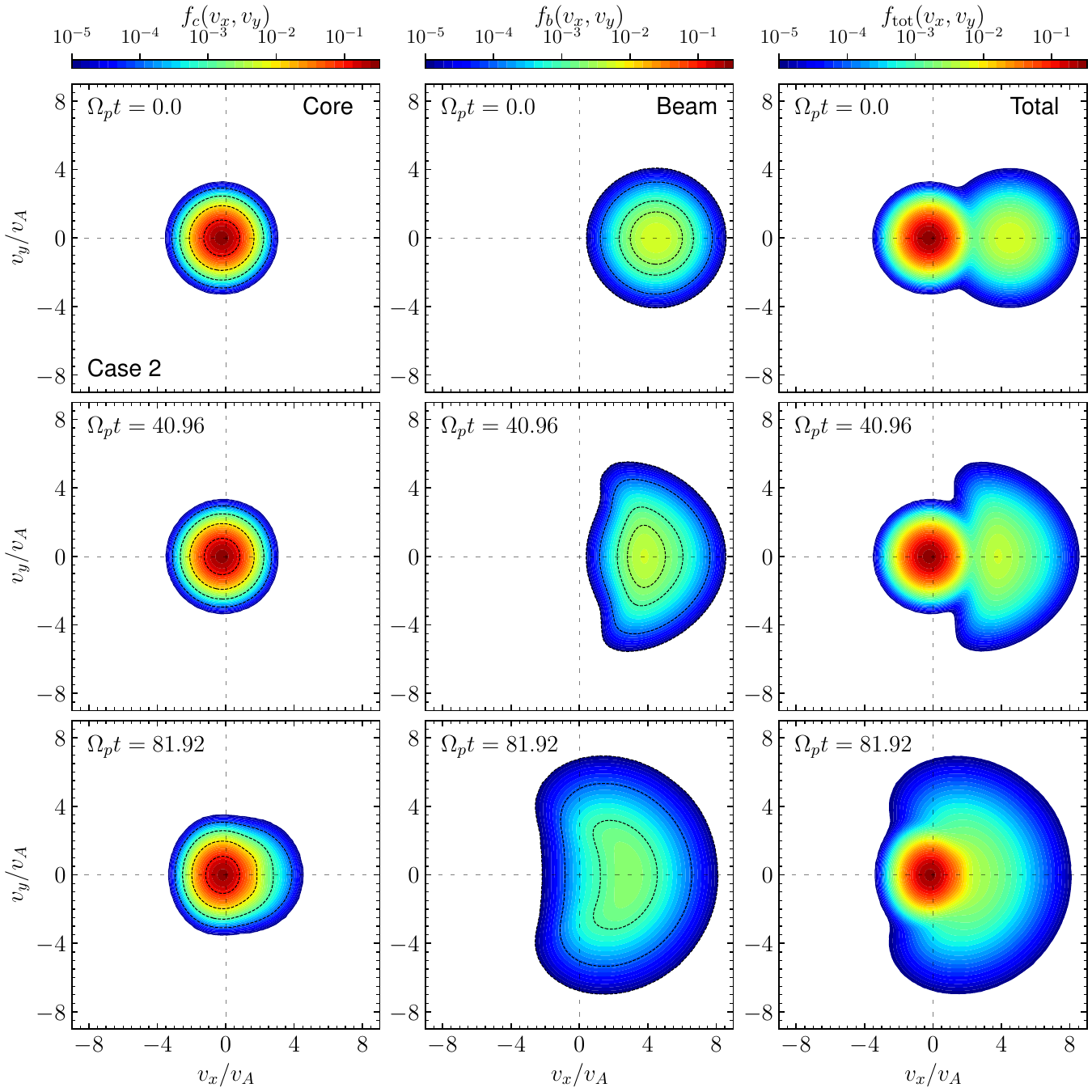}
\caption{Contours of VDs in the $v_x$-$v_y$ plane for various stages in the simulations for Case~1 (left) and Case~2 (right).}
        \label{fig:vdf}
    \end{figure*}
\end{center}
%
    \begin{figure}\centering
        \includegraphics[width=0.985\linewidth]{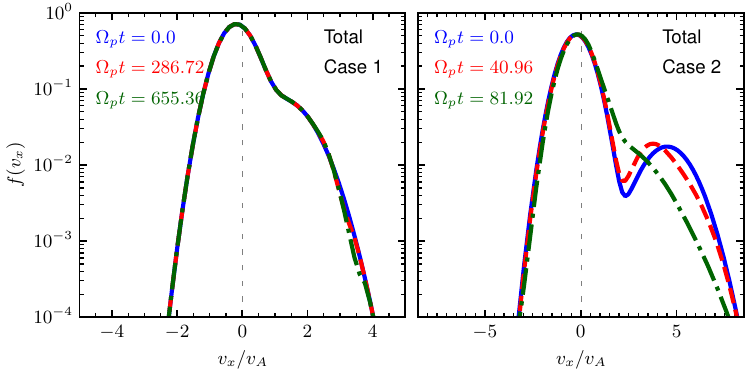}
        \caption{Parallel cuts of the VD function in the $v_x$, for various stages of both simulations. Top panels for Case 1, bottom panels for Case 2.}
        \label{fig:vdf1d}
    \end{figure}

Figure~\ref{fig:vdf} is divided into results for Case~1 (left) and Case~2 (right), showing snapshots of the proton VDs $f_p(v_x, v_y, t)$ as they evolve in time. Each division displays the core (left panels), beam (middle panels), and total (right panels) distributions, illustrating progressive deformations relative to their initial ($\Omega_p t=0$) shapes in response to the enhanced right-handed fluctuations. Apart from the initial condition, we include snapshots during the linear growth phase, and once the saturation is reached. For Case~1, the snapshots are taken at $\Omega_p t = 0$, $286.72$, and $655.36$, while for Case~2 they correspond to $\Omega_p t = 0$, $40.96$, and $81.92$. These represent the initial state, the phase of strong fluctuation growth, and the onset of wave–energy saturation shown in Figure~\ref{fig:moments}.  

To highlight the deformation of the proton components in the distribution, we have selected four particular iso-contours, as indicated with dotted lines, at $10^{-4}$, $10^{-3}$, $10^{-2}$, and $10^{-1}$. For case~1, one can see that the initial shape of the core distribution is generally preserved as time evolves, indicating weak wave--core proton interactions. In contrast, the initial beam distribution undergoes a modest
deformation, primarily through relaxation driven by diffusion in velocity space from the parallel direction ($v_x$) toward the perpendicular direction ($v_y$). This process corresponds to perpendicular heating of the beam population accompanied by a reduction in its drift speed. These deformations become increasingly pronounced as the system evolves, and by saturation ($\Omega_p t = 655.36$), the beam VD becomes more stable, retaining a reduced yet finite drift velocity and exhibiting a temperature anisotropy satisfying $A_b(0) < A_b(t) < 1$. The total distribution in the right panel clearly illustrates the overall deformation of the VD, which is primarily driven by beam-wave interaction. The core remains nearly unchanged, while beam-induced modifications give the combined distribution its final shape. The shapes of the total distributions show clear similarities to the results of the fully kinetic simulation in \citet{Pezzini-etal-2024}. Therefore, the hybrid model appears sufficient to capture the dominant processes responsible for this behavior.

\begin{center}
    \begin{figure}
        \centering
        \includegraphics[width=\linewidth]{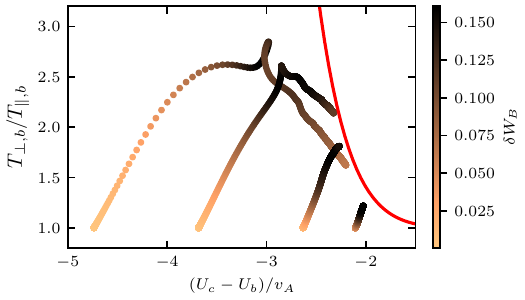}
        \caption{Dynamical path of various simulations with different initial relative drift, beam temperature anisotropy vs. relative drift. All four simulations use the input parameters of Table~\ref{t2} \citep{Verniero-etal-2022}, except for the drift.}
        \label{fig:paths}
    \end{figure}
\end{center}

In case~2, the deformations of the VD are significantly larger than in Case~1, consistent with the much stronger level of enhanced fluctuations, which drives a more effective wave-particle interaction involving both the core and beam populations. The initially isotropic proton core experiences perpendicular heating at early times ($\Omega_p t=40.96$), with particles diffusing along $v_y$ and developing a modest perpendicular temperature anisotropy $A_c>1$. As the system evolves (i.e., at $\Omega_p t=81.92$), the core then undergoes strong parallel heating, and particles diffuse toward the parallel direction, leading to a transition to a parallel temperature anisotropy $A_c<1$. Throughout this evolution, the core drift velocity is only slightly reduced.

The interaction of the enhanced fluctuations with the proton beam is even more pronounced. The initially isotropic beam undergoes rapid and strong deformation at early times ($\Omega_p t = 40.96$), exhibiting intense perpendicular heating and a significant reduction in the drift velocity. This leads to the development of a strong perpendicular anisotropy through diffusion of particles from the parallel ($v_x$) to the perpendicular
($v_y$) direction in velocity space. As time progresses (e.g., at $\Omega_p t = 81.92$), the beam drift continues to relax to lower values, and the large temperature anisotropy observed at early times gradually decreases, although it remains substantial by the end of the evolution. The combined VD, which incorporates the concurrent responses of both the core and the beam at the selected times, clearly displays the cumulative signature of these processes. At late stages, the resulting total distribution closely resembles the HH proton populations reported by \cite{Verniero-etal-2022}.

Further confirmation of the interpretations presented for the results in Figure~\ref{fig:vdf}, that the initial drift velocities are relaxed by the enhanced RH fluctuations and the beam component ended up with small but finite drift velocities, can be obtained by examining the parallel cuts of the VDs. Thus, in Figure~\ref{fig:vdf1d} we compare the parallel cuts of the VDs at different times. These 1-D cuts, $f_p(v_x, v_y = 0)$, of the total proton distribution reveal clear signatures of wave--particle resonant interaction. The temporal evolution of the VD in case~2 is much stronger than in case~1. In case~2, the beam population experiences a faster reduction in its drift velocity, as expected from the previous results, becoming evident already at $\Omega_p t = 81.91$, as shown by the dot-dashed green line.

A more suggestive representation of the relaxation of the drift velocity and the induced temperature anisotropy of the beam population, resulting from the wave-particle interaction between the beam protons and the enhanced RH fluctuations shown in Figure~\ref{fig:moments}, is presented in Figure~\ref{fig:paths}. There, we have run four simulations using the parameters in Table~\ref{t2} with different drift velocities, all until the same simulation time, $T_\text{max}=327.68/\Omega_p$. The dynamical paths in the anisotropy vs. relative drift, $(T_{\perp,b}/T_{\parallel,b},~(U_{c}-U_{b})/v_A)$-space, begin with initially isotropic beams $A_b(0)=1$. These trajectories show a clear reduction of the drift velocity accompanied by perpendicular heating, leading to pronounced induced temperature anisotropy $A_b(t)>1$ that, for example, can reach a value of about~$2.8$ before eventually relaxing to the quasistable state near $A_b \approx 1.55$ at the final stage; see the first trajectory from the left. 
We further directly compare the dynamical paths obtained from the hybrid simulations with the QL instability threshold proposed by \cite{Shaaban2024+AA}, shown by the red line. This threshold represents the saturated final states predicted by the QL analysis in the parameter space of beam--core drift velocity and beam temperature anisotropy, where the growth rates vanish at saturation, $\gamma_{\mathrm{m}}\rightarrow 0$. It is obtained from a parametric investigation that included variations of the initial drift velocity, beam density fraction, core parallel beta, and beam temperature anisotropy, namely $|U_{c}-U_b|/v_{A}\simeq 2.0$--$4.5$, $n_b/n_p=0.01$--$0.06$, $\beta_{\parallel c}=0.1$--$1.0$, and $A_b(0)=0.62$--$2.0$ \citep{Shaaban2024+AA}. The resulting marginal-stability boundary is fitted by a power-law function, following a standard practice in kinetic studies \citep{Gary1994,Shaaban2018+MNRAS}. Within the explored PSP-relevant range, all these parameter variations consistently relaxed toward the same threshold, indicating that the fitted boundary is not tied to a single initial condition but captures a robust relaxation path of the RH instability.

The final states from the hybrid simulations clearly lie along the QL instability threshold, with the unstable regimes situated to the right of the threshold at higher drift velocities. This agreement demonstrates consistency not only between the hybrid simulations and the QL predictions but also with the observed limits on the HH populations reported in \cite{Verniero-etal-2022}. Thus, the threshold provides a unified constraint linking the nonlinear hybrid evolution, the QL marginal-stability prediction, and the PSP observations. However, its extension to substantially larger beam fractions, such as $n_b/n_p\sim 0.1$ or a few tens of percent, remains to be tested.

\section{\bf Conclusions}\label{s4}

In this paper, we have investigated the excitation and long-term evolution of RH electromagnetic instabilities driven by proton beams, focusing on their relaxation to explain recent PSP observations in the young solar wind. 
The proposed parameterizations, as shown in Tables~\ref{t1} and \ref{t2}, and the results are consistent with PSP in situ data, capturing key proton properties and RH-polarized wave fluctuations \citep{Klein2021+ApJ, Verniero-etal-2022}.
Our 2D hybrid simulations, guided by linear and quasilinear theory \citep{Shaaban2024+AA}, demonstrate that the initial proton VD determines both the level of wave fluctuations and the subsequent deformation of the VDs.

The comparative linear analysis in Fig.~\ref{fig:linear} shows that Case 2, featuring higher drift speed and proton beta than Case 1, induces a significantly stronger instability.
Our hybrid simulations confirm these predictions: The magnetic wave energy in Case~2 grows by nearly two orders of magnitude from its initial state and saturates faster than in Case~1 (Fig.~\ref{fig:energy}), where the instability develops more slowly and peaks at a lower saturation level. 
In Case 1, the effect of RH fluctuations on the VDs, although limited, is mainly restricted to the beam population.
The core distribution preserves its initial profile throughout the simulation, while the beam exhibits modest transverse heating and a marginal decrease in drift velocity (Fig.~\ref{fig:vdf}). 
The corresponding moments in Fig.~\ref{fig:moments} show small changes in the beam anisotropy and nearly unchanged core parameters. 
In Case~2, the much stronger fluctuations lead to significant deformations of both proton populations. 
The core exhibits an initial perpendicular heating followed by enhanced parallel heating, leading to parallel anisotropy at later times (Fig.~\ref{fig:moments}, third row). 
The beam experiences faster and stronger deformations in this case, including intense perpendicular heating, significant drift relaxation (by more than 50\%), and large induced anisotropies (Fig.~\ref{fig:moments}, fourth row). 
These beam changes in the (total) VDs, as shown in the right panels of Fig.~\ref{fig:vdf}, closely resemble the HH profiles reported in PSP observations. 
The parallel cuts of the total VDs provided, as shown in Fig.~\ref{fig:vdf1d}, provide additional evidence for drift relaxation. 
In contrast, the evolution of VDs in Case~2 is significantly faster and more pronounced, correlating with enhanced wave fluctuations.

A key result of this study is presented in Fig.~\ref{fig:paths}. 
The dynamical paths of the beam parameters in the $(A_b,\,(U_{c}-U_{b})/v_A)$ space evolve towards the QL instability threshold derived by \citet{Shaaban2024+AA}. 
The final states of all simulations lie along this QL boundary, consistent with both theoretical predictions and the observed limits on HH proton populations reported in \citet{Verniero-etal-2022}. 
This alignment demonstrates that the nonlinear relaxation from the hybrid simulations converges to the same stability limit predicted by QL theory, providing strong justification for using the QL approach as a fast and reliable technique for analyzing such instabilities and the associated wave-particle interactions. It is also worth noting that the leftmost dynamical path in Figure~\ref{fig:paths} corresponds to the Case~2 results obtained using the plasma parameters listed in Table~\ref{t2}.

To conclude, our results show that proton beam–driven RH instability can play a central role in shaping the kinetic structure of proton VDs in the young solar wind.
Strong RH fluctuations, generated when sufficient beam energy is available, can naturally lead to the formation of HH distributions through self-consistent wave–particle interactions.
These findings bridge linear theory, nonlinear hybrid simulations, and recent PSP observations, offering a cohesive framework that also aligns with the QL analysis in \cite{Shaaban2024+AA}.

\section*{acknowledgments}

The authors thank Aaron Tran for helpful discussions on the details of the Hybrid-VPIC code.
We also acknowledge support from the Chilean Nuclear Energy Commission, Ruhr University Bochum, Katholieke Universiteit Leuven, and Qatar University. R.A.L.\ acknowledges the support of ANID Chile through FONDECyT grant No.\ 1251712. S.M.S. acknowledges the support of the Alexander von Humboldt Foundation through a short research stay grant, hosted at Ruhr University Bochum. SP is funded by the European Union (ERC-AdG agreement No 101141362, Open SESAME). L.P.\ acknowledges support from a PhD fellowship in fundamental research awarded by the Research Foundation Flanders (FWO), under grant number \href{https://app.dimensions.ai/details/grant/grant.13861380}{11PCB24N}. Views and opinions expressed are, however, those of the author(s) only and do not necessarily reflect those of the European Union or the European Research Council. Neither the European Union nor the granting authority can be held responsible for them. These results were also obtained in the framework of the projects C16/24/010 (C1 project Internal Funds KU Leuven), G0B5823N and G002523N (WEAVE) (FWO-Vlaanderen), and 4000145223 (SIDC Data Exploitation (SIDEX2), ESA Prodex). Powered@NLHPC: This research was partially supported by the supercomputing infrastructure of the NLHPC (CCSS210001). We also acknowledge the support of the International Space Science Institute (ISSI) in Bern, through ISSI International Team project 'Excitation and Dissipation of Kinetic-Scale Fluctuations in Space Plasmas' (ISSI Team project \#$24$-$612$).  Microsoft Copilot (\url{https://copilot.microsoft.com}) was used for proofreading and language refinement. We thank the anonymous reviewer for constructive criticism and suggestions.


\bibliography{papers}{}
\bibliographystyle{aasjournal}
\end{document}